# Identification of images of COVID-19 from Chest X-rays using Deep Learning: Comparing COGNEX VisionPro Deep Learning 1.0™ Software with Open Source Convolutional Neural Networks

Arjun Sarkar[1,2,*], Joerg Vandenhirtz, PhD[2], Jozsef Nagy[2], David Bacsa[2], Mitchell Riley[2]

1. FH Aachen University of Applied Sciences, Germany, Department of Biomedical Engineering
2. COGNEX Corporation, which funded the research in this paper.
*e-mail : arjun.sarkar@cognex.com

**ABSTRACT**

The novel Coronavirus – COVID-19 pandemic has been having a severe and catastrophic effect on humankind and is being considered the most crucial health calamity of the century. So far, one of the best methods of detecting COVID-19 has been from radiological images, namely X-Rays and Computed Tomography (CT) scan images. Many companies and educational organizations have come together during this crisis and created various Deep Learning models for the effective diagnosis of COVID-19 from chest radiography images. For example, The Centre for Bioengineering and Biotechnology from the University of Waterloo, along with Darwin AI – a startup spin-off of this department, has designed their Deep Learning model 'COVID-Net' and created a dataset called 'COVIDx' consisting of 13,975 images across 13,870 patient cases. In this study, COGNEX's Deep Learning Software: VisionPro Deep Learning™, which is used across various domains ranging from factory automation to life sciences, has been used to classify these Chest X-rays from the COVIDx dataset. The results are compared with the results of COVID-Net and various other state of the art Deep Learning models from the open-source community. Deep Learning tools are often referred to as black boxes because humans cannot interpret how or why a model is classifying an image into a particular class. This problem is addressed by testing VisionPro Deep Learning with two settings, firstly by selecting the entire image, that is, selecting the entire image as the Region of Interest (ROI), and secondly by segmenting the lungs in the first step, and then doing the classification step on the segmented lungs only, instead of using the entire image. VisionPro Deep Learning results: on the entire image as the ROI it achieves an overall F-score of 94.0%, and on the segmented lungs, it gets an F-score of 95.3%, which is at par or better than COVID-Net and other state of the art open-source Deep Learning models.

**INTRODUCTION**

The novel coronavirus disease – named COVID-19 by the World Health Organization– is caused by a new coronavirus class known as the SARS-CoV2 (Severe Acute Respiratory Syndrome Coronavirus 2). It is a single-stranded RNA (Ribonucleic Acid) virus that causes severe respiratory infections. The first COVID-19 cases were reported in December 2019, in Wuhan, Hubei province, China [1]. Since the virus has spread worldwide, it has been given the status of a pandemic by the World Health Organization. As of 30th July 2020, 12:00 GMT, 17.2 million people have been infected, and 670 thousand people have died due to COVID-19 [2]. There have been no vaccines available, so far, for treating COVID-19. One of the best solutions has been detecting the virus in its early stages and then isolating the infected people by quarantining them, thus preventing healthy people from getting infected.

In many cases, Real-time Reverse Transcriptase-Polymerase Chain Reaction (RRT-PCR) of nasopharyngeal swabs have been used for diagnosis [3]. The RT-PCR throat swabs are collected from patients with COVID-19, and the RNA is then extracted. This process takes over two hours to complete and has a long turnaround time with limited sensitivity. The best alternative is to detect images of COVID-19 from radiology scans [4,5,6] (chest X-ray images and chest Computed Tomography (CT) images). The advantages of using chest X-rays over CT images are as follows: X-ray imaging systems are much more widely available than CT imaging systems, they are cost-effective, and digital X-ray images can be analyzed at the same point of acquisition, thus making the diagnosis process extremely quick [7].

X-ray images are grayscale images. In medical imaging terms, these are images with values ranging from 0 to 255, where 0 corresponds to the completely dark pixels, and 255 corresponds to the completely white pixels. Different values on the X-ray image correlate to different areas of density. The different values are - Dark: Locations in the body which are filled with air are going to appear black, Dark Grey: Subcutaneous tissues or fat, Light Grey: Soft tissues like the heart and blood vessels, Off White: Bones such as the ribs, Bright White: Presence of metallic objects such as pacemakers or defibrillators. The way physicians interpret an image is by looking at the borders between the different densities. The ribs appear off-white because they are dense tissues, but since the lungs are filled with air, the lungs appear dark. Similarly, below the lung is the hemidiaphragm, which is a soft tissue and hence appears light grey. This helps in finding the location and extent of the lungs. If two objects with different densities are present close to each other, they can be demarcated in an X-ray image. If something happens in the lungs, such as Pneumonia, the air-dense lungs change into water-dense lungs. This causes the demarcation lines to fade since the pixel densities start closing in on the grayscale bar [8].

About 20% of the patients infected with COVID-19 develop pulmonary infiltrates and some develop very serious abnormalities [9]. The virus reaches the lungs' gas exchange units and infects alveolar type 2 cells [10][11]. The most frequent CT abnormalities observed are ground-glass opacity, consolidation, and interlobular septal thickening in both lungs [12]. But due to infection control issues related to patient transport to CT rooms, problems encountered in CT room decontamination, and the lack of CT scanners availability in different parts of the world, portable chest X-rays are likely to be one of the most common modalities for identification and follow up of COVID-19 lung abnormalities [13]. Hence a significant number of expert radiologists who can interpret these radiology images are needed. Due to the ever-increasing number of cases of COVID-19 infections, it is getting harder for radiologists to keep up with this demand. In this scenario, Deep Learning techniques prove to be beneficial in both classifying the abnormalities from lung X-ray images and in aiding the radiologists to accurately predict COVID-19 cases in a reduced time frame.

While many studies have demonstrated success in detecting images of COVID-19 using Deep Learning with both CT scans and X-rays, most of the Deep Learning architectures need extensive programming . Moreover, most of the architectures fail to showcase if the Deep Learning model is being triggered by abnormalities in the lungs or on some artifacts not related to COVID-19. Due to the absence of a GUI (Graphical User Interface) with most of these Deep Learning models, it is difficult for radiologists, who lack knowledge in Deep Learning or programming, to use these models, let alone train them. Therefore, we showcase an already existing Deep Learning software with a very intuitive GUI, which can be used as a pretrained software or can even be trained on new data from particular hospitals or research centers.

COGNEX VisionPro Deep Learning[TM] is a Deep Learning vision software, from COGNEX Corporation (Headquarters: Natick, MA, United States). It is a field-tested, optimized, and reliable software solution based on a state-of-the-art set of machine learning algorithms. VisionPro Deep Learning combines a comprehensive machine vision tool library with advanced Deep Learning tools.

In this study, we used the latest version - VisionPro Deep Learning 1.0 to aid in the classification of images as Normal, Non COVID-19 (Pneumonia), or COVID-19 chest X-rays. The results are compared with various state of the art open-source neural networks.

The VisionPro Deep Learning GUI, also called COGNEX Deep Learning Studio, has three tools for image classification, segmentation, and location. It contains various Deep Learning architectures built within the GUI, to carry out specific tasks:

1) Green Tool – This is the Classify tool. It is used to classify objects or complete scenes. It can be used to classify defects, cell types, images of different labels, or different types of test tubes used in laboratories. The Green tool learns from the collection of labeled images of different classes and can then be used to classify images that it has not seen previously. This tool is similar to classification neural networks such as VGG [14], ResNet [15] or DenseNet [16].

2) Red Tool – This is the Analyze tool. It is used for segmentation and defect/anomaly detection, for example, to aid in the detection of anomalies in blood samples (clots), incomplete or improper centrifugation, or sample quality management. The Red tool is also used to segment specific regions, such as defects or areas of interest. The Red tool comes with the option of using either Supervised Learning or Unsupervised Learning for segmentation and detection. This is similar to the segmentation neural network, such as U-Net [17].

3) Blue Tool –

 a) This is the Feature Localization and Identification tool. The Blue tool finds complex features and objects by learning from labeled images. It has self-learning algorithms that can locate, classify, and count the objects in an image. It can be used for locating organs in X-ray images or cells on a microscopic slide.

 b) The Blue tool also has Read feature. It is a pretrained model that helps decipher severely deformed and poorly etched words and codes using optical character recognition (OCR). This is the only pretrained tool. All the other tools need to be trained on images first to get results.

For classification of COVID-19 images, two settings are used:

1) Green tool for classification of the entire chest X-ray images

2) Red tool for segmentation of the lungs, and then, a subsequent Green tool classifier just running on the segmented lungs to make sure the Deep Learning software predicts its results based on just the lungs.

**METHOD**

**1. Dataset**

The open access benchmark dataset called COVIDx is used for training the various models [18]. The dataset contains a total of 13,975 Chest X-ray images from 13,870 patients. The dataset is a combination of five different publicly available datasets. According to the authors [18] of COVID-Net, COVIDx is one of the largest open source benchmark datasets in terms of the number of COVID-19 positive patient cases.

These five datasets were used by the authors of COVID-Net to generate the final COVIDx dataset:

a) Non-COVID 19 pneumonia patient cases and COVID-19 cases from the COVID-19 Image Data Collection [19],

b) COVID-19 patient cases from the Figure 1 COVID-19 Chest Xray Dataset [20],

c) COVID-19 patient cases from the ActualMed COVID-19 Chest X-ray Dataset [21],

d) Patient cases who have no pneumonia (that is, normal) and non-COVID-19 pneumonia patient cases from RSNA Pneumonia Detection Challenge dataset [22],

e) COVID-19 patient cases from COVID-19 radiography dataset [23].

The idea behind using these five datasets was that these are all open source COVID-19/Pneumonia Chest X-ray datasets, so they can be accessed by everyone in the research community and by the general public, and also add variety to the dataset. However, the lack of COVID-19 Chest X-ray images made the dataset highly imbalanced. Of the total 13,975 images, the data was split into 13,675 training images and the remaining 300 into test images. The data was divided across three classes, 1. Normal (for X-rays which did not contain Pneumonia or COVID-19), 2. Non-COVID-19/Pneumonia (for X-rays, which had some form of bacterial or viral pneumonia, but not COVID-19), and 3. COVID-19 (for X-rays which were COVID-19 positive). In the training set, there were 13,675 images, with 7966 of those belonging to the Normal class, 5451 images belonging to the class Non-COVID-19/Pneumonia, and only 258 images in the class COVID-19. The test set was a balanced set, with each of the three classes having 100 images each [18].

The authors of COVID-Net have shared the dataset generating scripts, for the construction of the COVIDx dataset for public access available at the following link - https://github.com/lindawangg/COVID-Net [18]. The python notebook 'create_COVIDx_v3.ipynb' is used to generate the dataset. The text files 'train_COVIDx3.txt' and 'test_COVIDx3.txt' contains the file names used in the training and test set, respectively. It is then tested with VisionPro Deep Learning, and the results are compared with COVID-Net results and other open-source Convolutional Neural Network (CNN) architectures such as VGG [14] and DenseNet [16]. Tensorflow [34] (developed by Google Brain Team [35]) library is used to generate and train the open source CNN architectures.

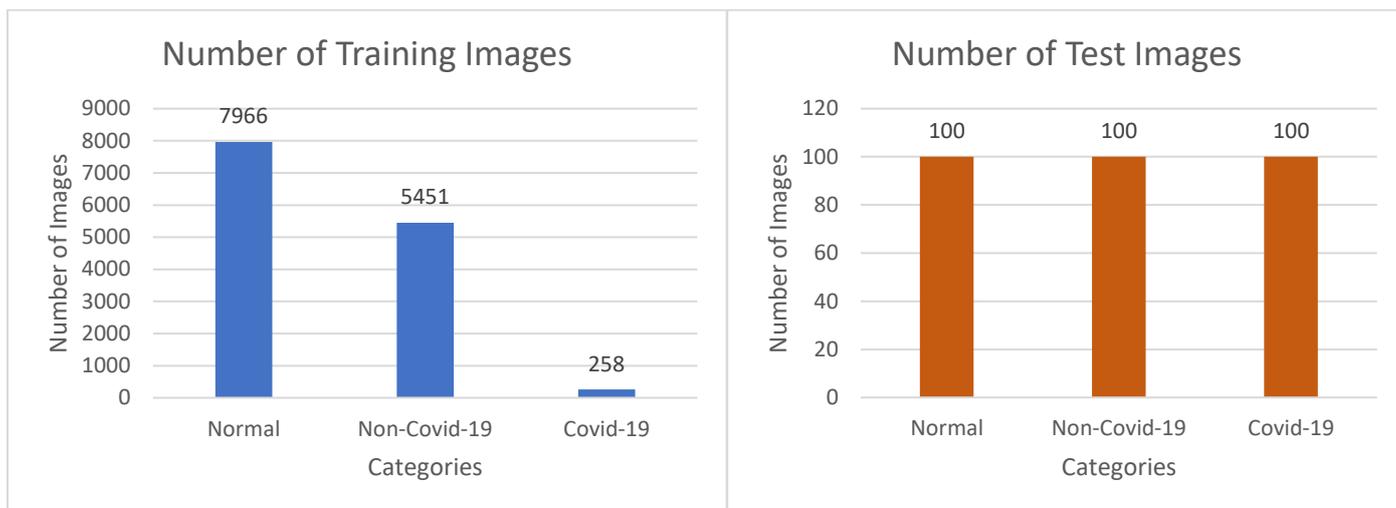

**Figure 1:** Chest X-ray image distribution of each class: Normal (no infection), Non-COVID-19 (Pneumonia), and COVID-19 images. In the training set, there are 7966 images belonging to class 'Normal', 5451 images in class 'Non-COVID-19', and 258 'COVID-19' images. In the test set, there is an equal distribution of 100 images across all the three classes. Horizontal axis represents the different categories or classes, and the vertical axis represents the number of images in each of these categories.

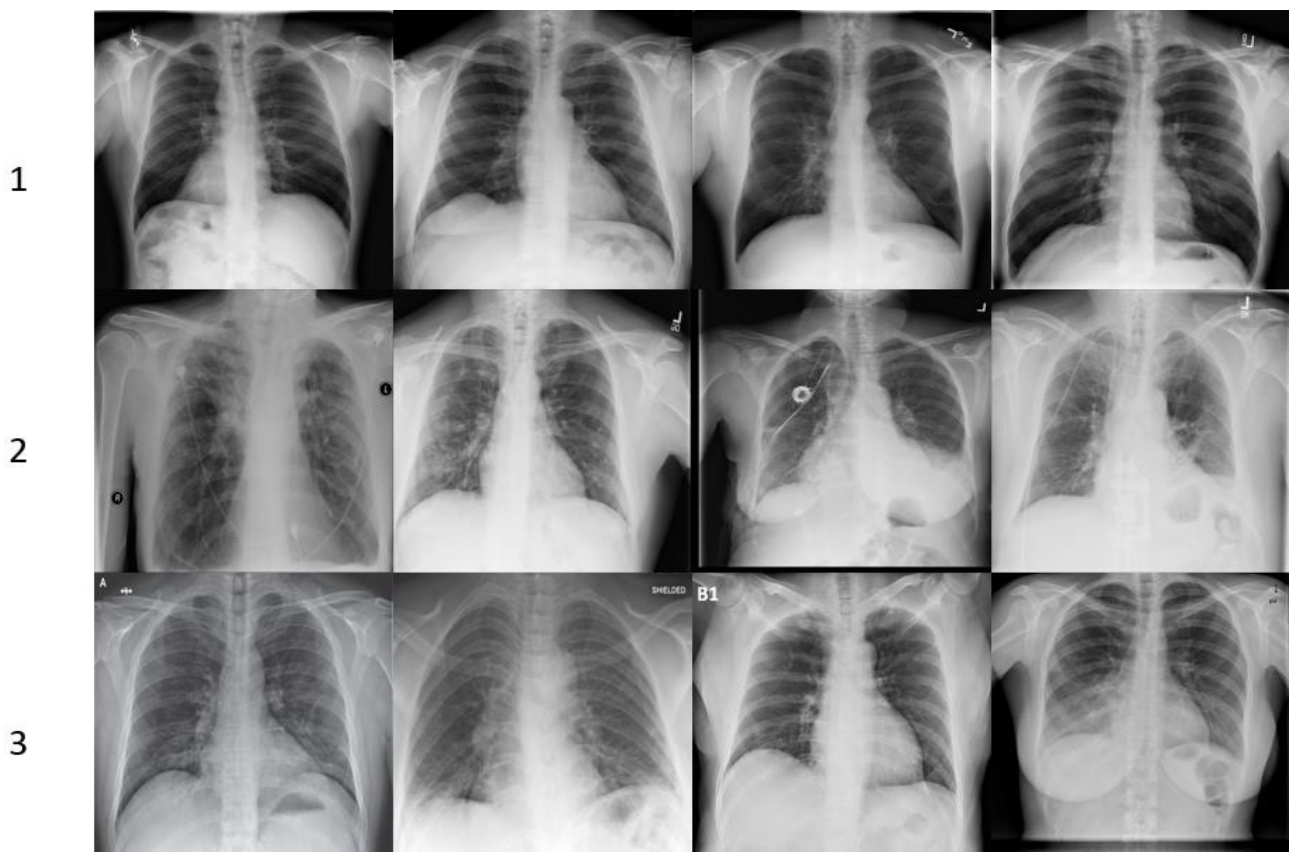

**Figure 2:** Examples of the chest X-ray images belonging to the different classes. The class numbers are shown along the vertical axis. Class 1: Normal images, Class 2: Non- COVID-19 (Pneumonia) images and Class 3: COVID-19 images. All the images belong to the training set of the COVIDx dataset [18].

## 2. Preprocessing Data

The scripts for generating the COVIDx dataset are used to merge the five datasets together and separate the images into train and test folders. Along with the images, the script also generates two text files containing the names of images belonging to the train and test folders, and their class labels [18].

To simplify classification, a python script is used to convert the '.txt' files into 'pandas' data frames and then finally converted to '.csv' files for better understanding. Next, another python script is created to rename all the X-ray images of the train and test folders according to their class labels and store them in a new train and test directories. Since the goal is classification of the X-ray images, renaming the images makes it easier to interpret the images directly from their file names, rather than consulting a '.csv' file every time. Finally, we have all the 13,975 images in train and test folders, with their file names containing the class labels.

### a) COGNEX VisionPro DEEP LEARNING Suite 5.0

Unlike most other Deep Learning architectures VisionPro Deep Learning does not need any preprocessing of the images. The images can be fed directly into the GUI, and the software automatically does the preprocessing, before starting to train the model.

Since the COVIDx dataset is a combination of various datasets, the images have different color depth, and VisionPro Deep Learning GUI found 326 anomalous images. Training could have been done keeping the anomalous images in the dataset, but it might have reduced the overall F-score of the model. Therefore, we normalize the color depth of all COVIDx images to 24-bit color depth using an external software, IrfanView (open source: irfanview.com). Then the images are added into the VisionPro Deep Learning GUI.

No other preprocessing steps are necessary with VisionPro Deep learning, such as - image augmentation or setting class weights or oversampling of the imbalanced classes, which are necessary for training the other open-source CNN models. Once the images are fed into the VisionPro Deep learning GUI, they are ready to be trained.

### b) Open Source Convolutional Neural Network (CNN) Models

Before training the CNN models, such as VGG [14] or DenseNet [16], it is necessary to execute some preprocessing steps, such as resizing, artificial oversampling of the classes with fewer images, image standardization and finally data augmentation. First the images are resized to 256x256 pixels. The entire training is done on a Nvidia 2080 GPU, and as to not run into 'GPU memory errors' this is found to be the perfect image size. Once the images are resized, and the images and labels loaded together, it is necessary to oversample images which belong to the classes having fewer images, that is, for Non-COVID-19 and the COVID-19 classes. For oversampling, random artificial augmentations are carried out, such as rotation (-20 degrees to +20 degrees), translation, horizontal flip, gaussian blur, and adding external noise. All these are applied randomly using the 'random' library in python. Then all the X-ray images are standardized to have values with a mean of zero and a standard deviation of 1. This is done, keeping in mind that standardization helps the Deep Learning network to learn much faster. Finally, data augmentation is added on all the classes, irrespective of the number of images belonging to those classes. Augmentations include rescaling, height and width shifting, rotating, shearing and zooming. After all these preprocessing steps, the images are ready to be fed into the deep neural networks.

## 3. Classification using VisionPro Deep Learning

The goal of the study is the classification of Normal, Non-COVID-19(Pneumonia) and COVID-19 X-ray images. For classification, VisionPro Deep Learning uses the Green tool. Once the images are loaded and labeled, they are ready for training. In VisionPro Deep Learning, the Region of Interest (ROI) of the images can be selected. Thus, it is possible to reduce the edges by 10-20% to remove artifacts like the letters or borders, which are usually at the edges of the images. In this case, the entire images are used without cropping the edges because many images have the lungs towards the edges, and we didn't want to remove essential information.

For feeding the images into VisionPro Deep learning, the images do not need to be resized. Images of all resolutions and aspect ratio can be fed into the GUI, and the GUI does the preprocessing automatically before starting the training.

In VisionPro Deep learning, the Green tool has two subcategories, High-detail and Focused. Under High-detail there are several options such as sizes of model architectures -Small, Normal, Large and Extra-Large models, which can be selected for training the model. We train the network using the High detail subcategory and selecting the 'Normal' size model.

Out of the 13,675 images, 80% of the images are used for training. The VisionPro Deep Learning suite automatically selects the other 20% images for validation. Both the training and validation sets are randomly selected by the VisionPro Deep Learning suite. The user just needs to specify the train-validation split. The maximum number of epoch counts are selected to be 100. There are options of selecting the minimum epochs and patience for which the model will train, but this is not selected. Once these are selected, training is started by clicking on the 'brain' icon on the green tool, as seen in Figure 3.

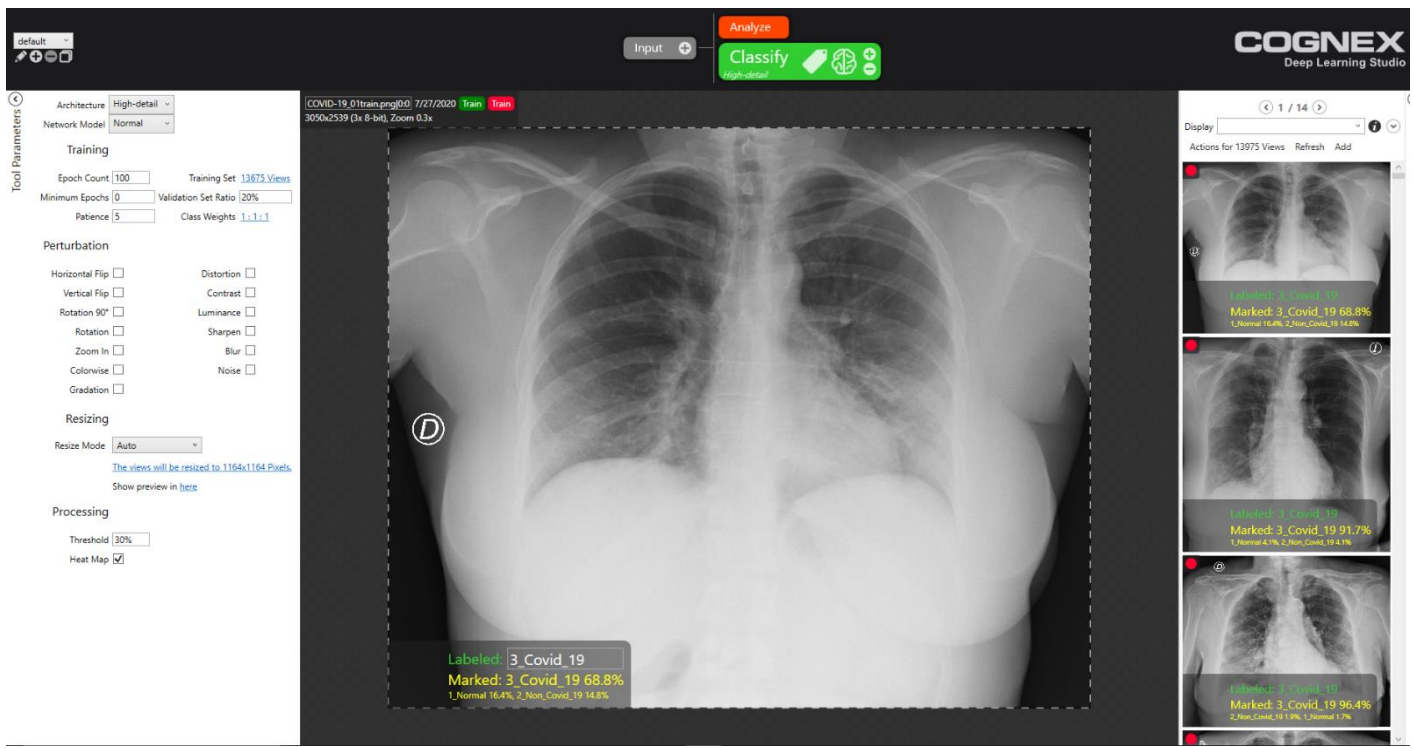

**Figure 3:** The VisionPro Deep Learning GUI loaded with the X-ray images from the COVIDx dataset [18]. On the left of the GUI, there are options to select various parameters for training the model, such as model type, model size, epoch count, minimum epochs and patience, train and validation split, class weights, threshold, heat-map and the different data augmentation options of flip, rotation, contrast, zoom, brightness, sharpen, blur, distortion and noise. In the middle, the selected image is shown. On the right, thumbnails of all the images in the training and test set are shown. On the top, there is the tool selection option. In the figure, the green tool has been selected for classification. Clicking on the 'brain' shaped icon in the green tool, starts the training of the model.

4. Segmentation and improved classification using VisionPro Deep Learning

The Green tool is used to classify entire X-ray images but for identification of images of COVID-19 the Deep Learning model needs to focus on the lungs, and not the peripheral bones, organs and soft tissues. The model must make its predictions exclusively based on the lungs and not on the differences in spinous process, clavicles, soft tissues, ornaments worn around the patient's neck or even the background. This way we can be sure that the model is classifying based entirely on the normal and infected lungs. Therefore, segmentation of the lungs from each image makes sure that the model trains only on these segmented lungs, and not on the entire image. To implement this, the VisionPro Deep Learning Red tool is used. The Red tool is used for segmenting the images, such that only the lungs are visible to the Deep Learning model for training. For achieving this, 100 images of the training set are manually masked using the 'Region selection' option in the Red Tool. The training set consists of 13,675 images, but a manual masking of 100 images is enough to train the model. Once the manual masking is done on the 100 images, the Red tool is trained. After training is completed, the VisionPro Deep Learning GUI has all the training and test images properly masked, such that only the lungs are visible. Anything outside the lungs is treated as outside the ROI and is not used in classification. The Red tool is added in the same environment as the previous green tool and there is no need to create a new instance for segmentation.

Once all the images are segmented, a Green classification tool is implemented after the Red tool. The Green tool is then used to start the classification (similar to Step 3 of Methods), but this time exclusively on the segmented lungs and not on the entire images.

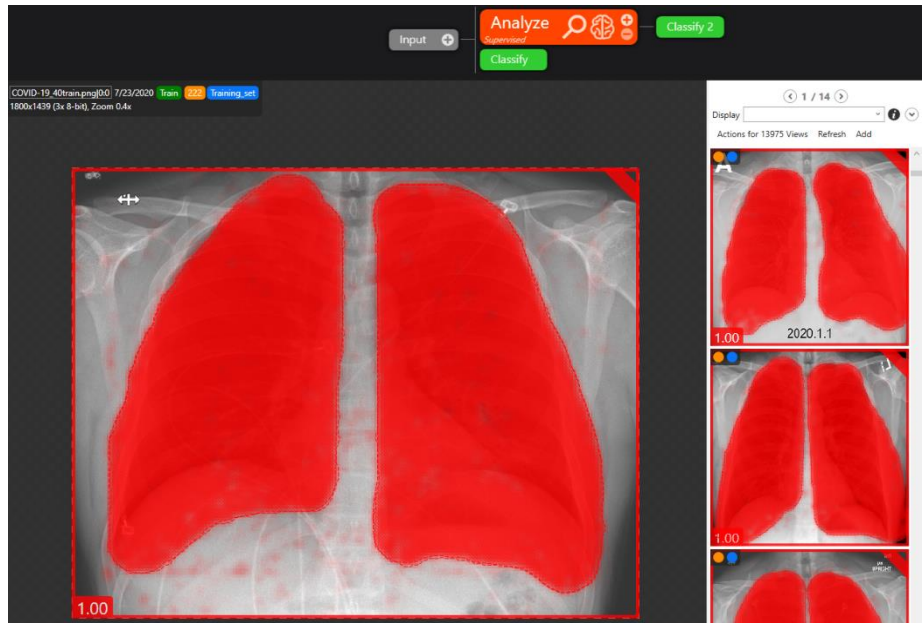

**Figure 4**: Lungs masked using the Red Tool. 100 such images are manually masked. Then the Red tool is trained. This helps to mask all the images in the training set and are used later for classification.

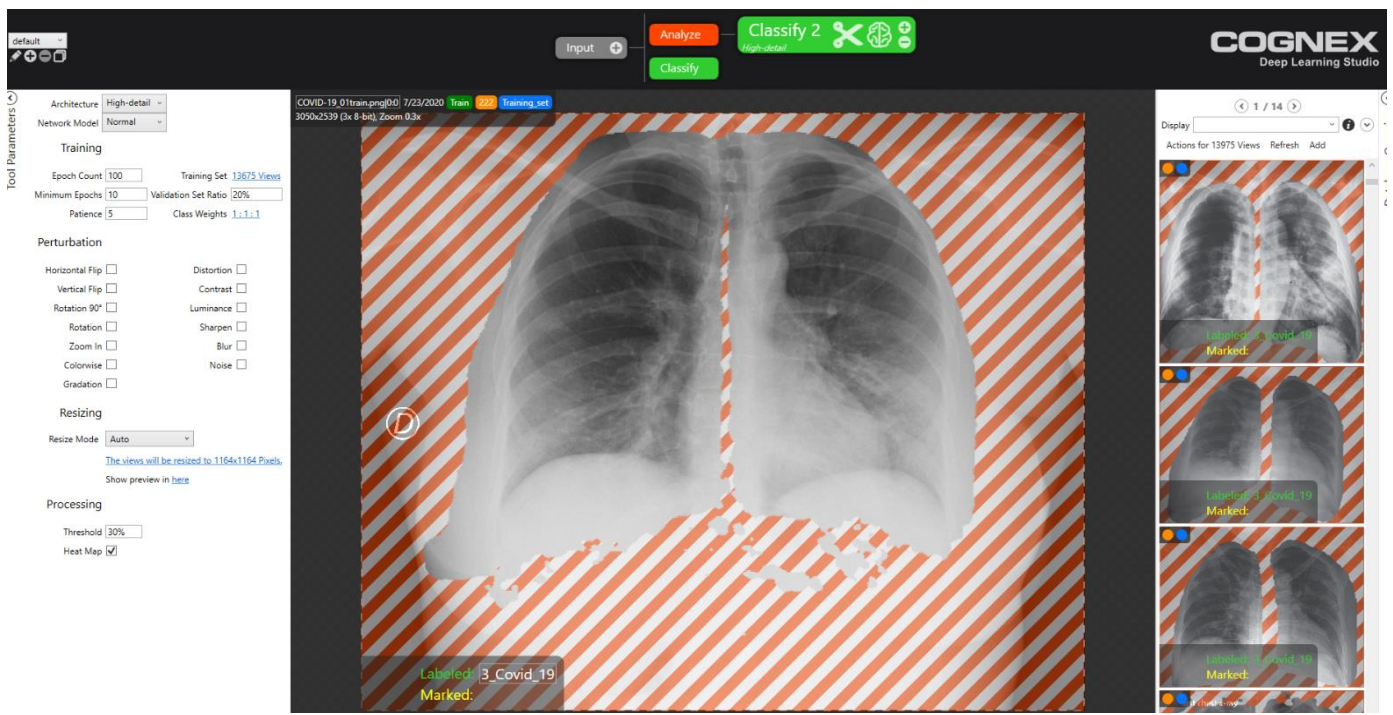

**Figure 5:** The segmented lungs after training the Red tool in VisionPro Deep Learning. Anything outside the segmented lungs is not considered to be part of the Region of Interest (ROI) and is not used for classification. This makes sure that VisionPro Deep Learning trains only on the lungs and not on the artifacts around it.

## 5. Classification using VGG network

The VGG [14] network is a deep neural network and is still one of the state-of-the-art Deep Learning models used in image classification.

We use the 19-layer VGG 19 model for training using transfer learning, on the COVIDx dataset. VGG takes an image of size 224x224 pixels. Preprocessing of the images are done automatically by calling 'preprocess input' from the VGG19 model in TensorFlow. The 'preprocess input' is fed into the 'ImageDataGenerator' from TensorFlow(Keras). 'ImageNet' weights are used for training. The COVIDx dataset is also resampled as stated in 2 (b) of the methods section. This ensures that all classes have similar number of images, so as to avoid the model favoring a particular class during training. The VGG19 architecture uses 3x3 convolutional filters which performs much better than the older AlexNet [24] models. All the activation functions used in the hidden layers are ReLU (Rectified Linear Units) [25]. After the VGG architecture we add four fully connected layers with 1024 nodes each. All four layers use the ReLU activation function and L2 regularization [26][27]. To provide better regularization, after each of these layers a Dropout is set. The final layer is a fully connected layer of 3 nodes for the classification of the 3 classes. The final layer has the activation function 'SoftMax' [28].

In the preprocessing steps, the labels of the images are not 'one hot encoded' but kept as three distinct digits. So instead of using 'categorical cross entropy' [29], which is commonly used when the labels are one-hot encoded, the 'sparse categorical cross entropy' is used as the loss function. 'Adam' [30] optimizer is used with learning rate scheduling, such that the learning rate decreases after every thirty epochs. During training, several callbacks are set, such as saving the model each time the validation loss decreases, and using early stopping, to stop the training when the validation loss does not improve even after several epochs. The epoch count is set to 100. For training, batches of 32 images are fed to the model at once. Once all these hyperparameters are set, training of the model is started. After training completion, the program is set to plot the confusion matrix and give results on the various evaluation metrics, based on which the various models are compared.

## 6. Classification using ResNet

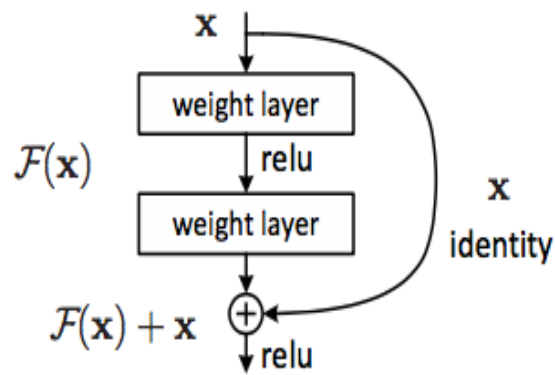

**Figure 6:** Residual learning: a building block. Image from original ResNet paper [15].

One of the bottlenecks of the VGG network is that it does not go too deep as it starts losing generalization capability the deeper it goes. To overcome this problem ResNet or Residual Network [15] is chosen.

The ResNet architecture consists of several residual blocks with each block having several convolutional operations. The implementation of skip connections, as shown in Figure 5, makes the ResNet better than VGG. The skip connections between layers add the outputs from previous layers to the outputs of the stacked layers. This allows the training of deeper networks. One of the problems that ResNet solves is the vanishing gradient problem [31].

For training the COVIDx dataset we use the 50-layer ResNet50V2 (Version 2) architecture. We use transfer learning to train the model, and then add eight fully connected layers with L2 regularization followed by Dropouts for better regularization. All the other settings and hyperparameters are kept similar to the training of the VGG19 network (Method, part 5).

## 7. Classification using DenseNet

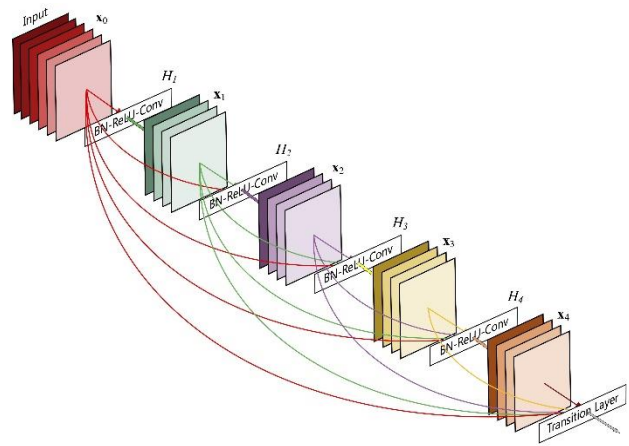

**Figure 7:** A 5-layer dense block. Each layer takes all preceding feature-maps as input. Image from original DenseNet paper [16].

DenseNet (Dense Convolutional Network) [16] is an architecture which focuses on making the Deep Learning networks go even deeper, but at the same time makes them more efficient to train, by using shorter connection between the layers. DenseNet is a convolutional neural network where each layer is connected to all other layers that are deeper in the network, that is, the first layer is connected to the 2nd, 3rd, 4th and so on, the second layer is connected to the 3rd, 4th, 5th and so on. Unlike ResNet [15] it does not combine features through summation but combines the features by concatenating them. So, the 'i$^{th}$' layer has 'i' inputs and consists of feature maps of all its preceding convolutional blocks. It hence requires fewer parameters than traditional convolutional neural networks.

For training the COVIDx dataset we use the 121 layered DenseNet121 architecture. We use transfer learning to train the model, and then add eight fully connected layers with L2 regularization followed by Dropouts for better regularization. All the other settings and hyperparameters are kept similar to the training of the VGG19 network (Method, part 5).

## 8. Classification using Inception Network

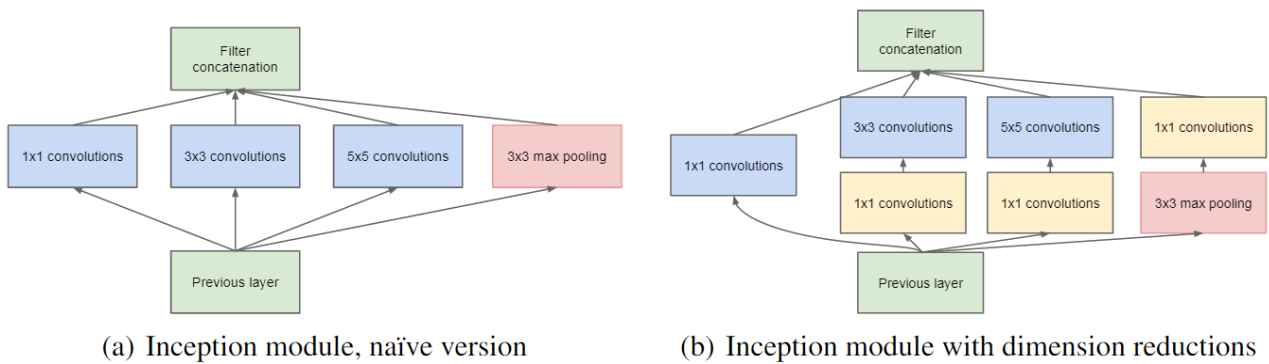

**Figure 8:** Inception Module as shown in the original Inception Version 1 paper [32].

Inception Network [32] has been developed with the idea of going even deeper with convolutional blocks. Very deep networks are prone to overfitting, and it is hard to pass gradient updates throughout the entire network. Also, images may have huge variations, and thus choosing the right kernel size for convolution layers is hard. To address these problems Inception network is one of the best possible networks. Inception network version 1 has multiple sizes of filters in the same level. It has various connections of 3 different sizes of filters of 1x1, 3x3, 5x5, with max pooling in a single inception module. All the outputs are concatenated and then sent to the next inception module.

For training the COVIDx dataset we use the 48 layered InceptionV3 [33] architecture, which also includes 7x7 convolutions, Batch Normalization and Label smoothing in addition to the Inception version 1 modules. We use transfer learning to train the model, and then add eight fully connected layers with L2 regularization followed by Dropouts for better regularization. All the other settings and hyperparameters are kept similar to the training of the VGG19 network (Method, part 5).

## RESULTS

### 1. Evaluation Metrics

In medical imaging, since the decisions are of high impact, it is very important to understand exactly which evaluation metrics are necessary to decide if a model works on a patient or not. Accuracy of a model is not the best metric for deciding whether the model is fit for a patient. Rather it is important to look into other evaluation metrices such as sensitivity, predictive values and overall F-scores.

First, the **confusion matrix** is plotted for the 300 test images, for all the models that we use for the comparison.

Figure 9 shows the confusion matrix of VisionPro Deep Learning on the entire ROI. VisionPro Deep learning GUI does not display numbers of correctly classified or misclassified images on the confusion matrix, but if any point on the confusion matrix is clicked, it displays, not only the number of images in that category, but also all the images belonging to that category, with the prediction percentage and whether the prediction it made is correct or not. Below the confusion matrix it displays all the evaluation metrics of recall (sensitivity), precision (positive predictive value) and F-scores. This table contains the number of labeled images, which shows the number of images in each class in the test set. The 'Found' column shows the number of images VisionPro Deep Learning thinks should belong in those classes.

A report can also be generated on all the test images as seen in Figure 10. It shows a small snippet of 6 COVID-19 positive images from the test set. The report contains details of the 300 test images, including the filename, the image, the original label as 'Labeled', and the predictions made by VisionPro Deep learning as 'Marked', with the percentage of confidence of prediction on each class. If the prediction is different from that of the label, then it is marked in red.

**Misclassification Results:** Out of the 300 test images, VisionPro Deep Learning classifies 18 images incorrectly with entire ROI selected, and 16 images incorrectly with the segmented lungs. COVID-Net had classified 20 images incorrectly [18]. VGG19 [14], ResNet50V2 [15], Densenet121 [16], and InceptionV3 [33] networks make 47, 37, 41, 26 misclassifications, respectively. **VisionPro Deep Learning has fewer misclassifications than all the open-source models in both the settings.** Compared to COVID-Net [18], the performance of VisionPro Deep learning is similar with the entire image as ROI and much better when using the segmented lungs.

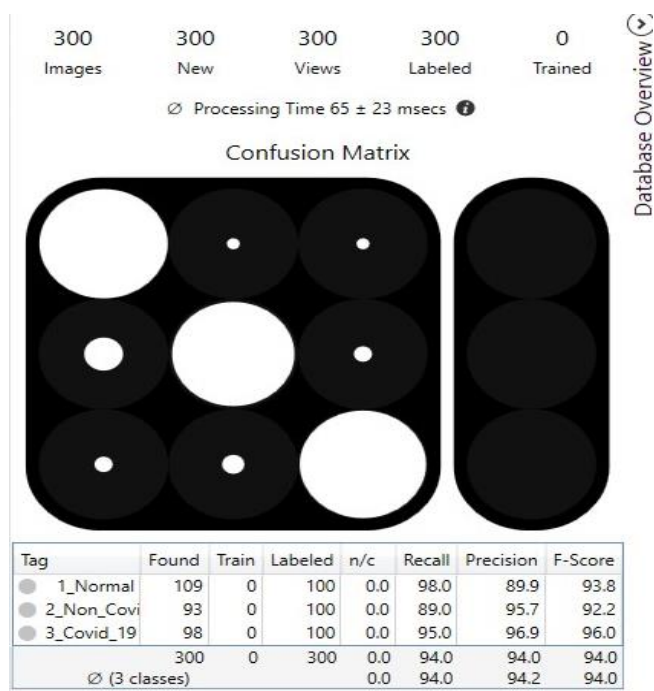

**Figure 9:** Confusion matrix on the 300 test images COGNEX VisionPro Deep Learning with entire ROI selected.

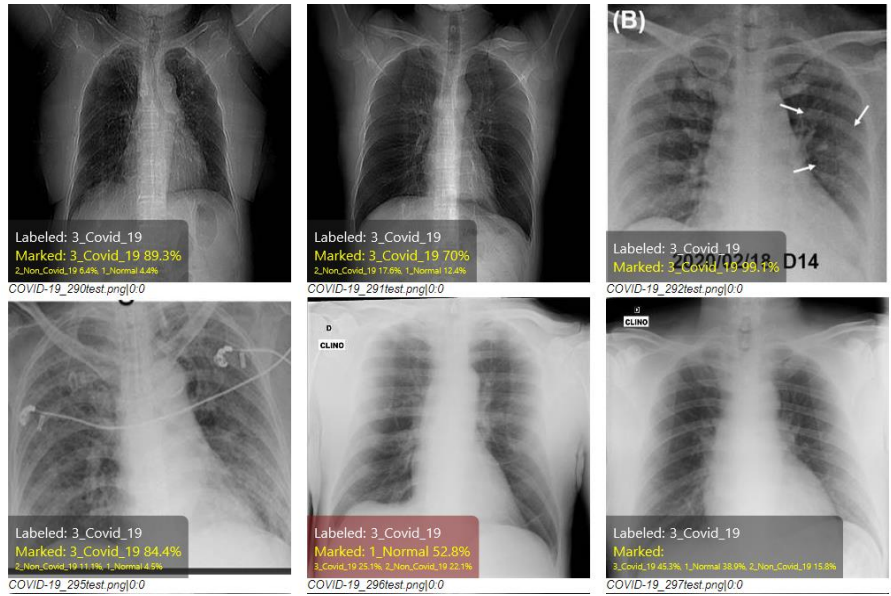

**Figure 10:** A snippet of the report generated on the 300 test images by VisionPro Deep Learning with the entire image selected as the ROI. The report contains the confusion matrix with the evaluation metrics: Sensitivity (Recall), Positive Predictive Value (Precision) and F-score for each class. the test images are also shown with the correct labels, the predicted labels and the confidence percentage of each class. In this image, 5 images are classified correctly, and 1 image misclassified (marked in red).

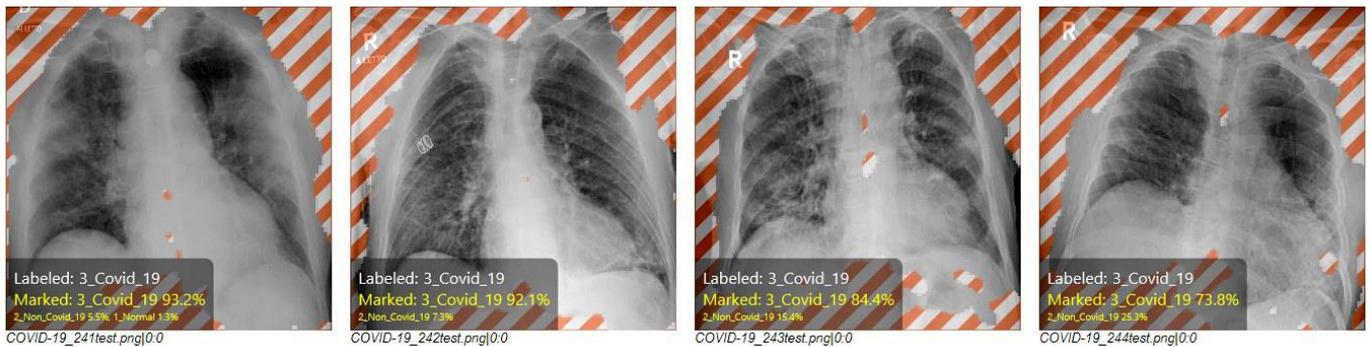

**Figure 11:** A snippet of the report generated on the 300 test images by VisionPro Deep Learning with the segmented lungs as the ROI. In this image, all 4 images are classified correctly.

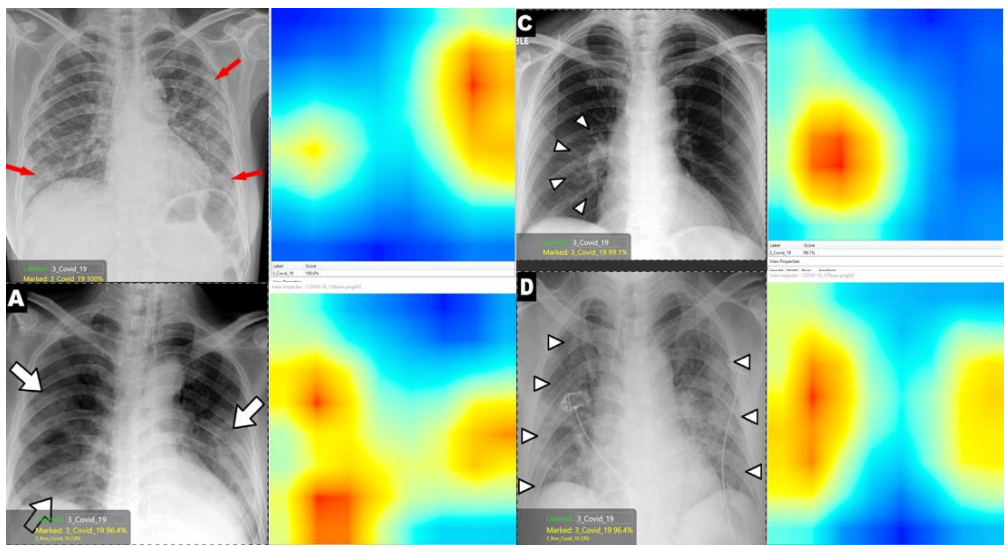

**Figure 12:** Four COVID-19 X-ray images from the test set of the COVIDx dataset along with the predicted heatmaps generated by VisionPro Deep Learning. As seen from the heatmaps, VisionPro Deep Learning makes its predictions based on the actual abnormalities and not on artifacts, as indicated by the arrows on the X-ray images.

**Heatmaps** are a great way to visualize predictions of the Deep Learning algorithm. They point out exactly which parts of the image triggers the model to come up with its predictions. Figure 12 shows the heatmaps generated by VisionPro Deep Learning on four COVID-19 images. The arrows on the X-ray images are indicating the position of the most infected parts of the lungs. The heatmaps clearly show, that VisionPro Deep learning generates its results based on exactly these positions on the lungs, indicating that the predictions are not based on artifacts, but rather on the actual abnormalities.

Among the open-source network architectures, the InceptionV3 gives the best results, followed by ResnetV2 [15]. The confusion matrix for all the open-source architectures and COVID-Net [18] are shown in Figure 13,14, respectively.

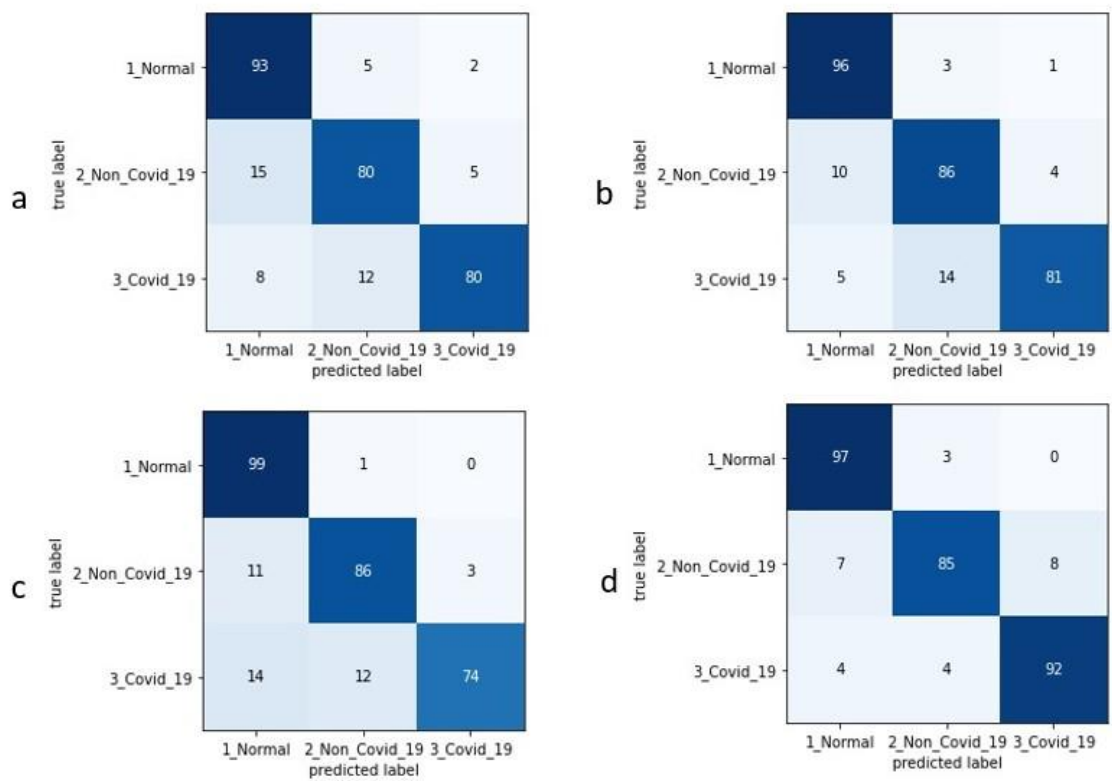

**Figure 13:** Confusion matrix on the 300 test images for the open-source architectures a: VGG19 [14], **b**: Resnet50 V2 [15], **c**: Densenet121 [16], **d**: Inception V3 [33]. Inception V3 has the best results, with the least number of false predictions. ResNet50 V2 has the next best result, followed by DenseNet121 and VGG19, respectively.

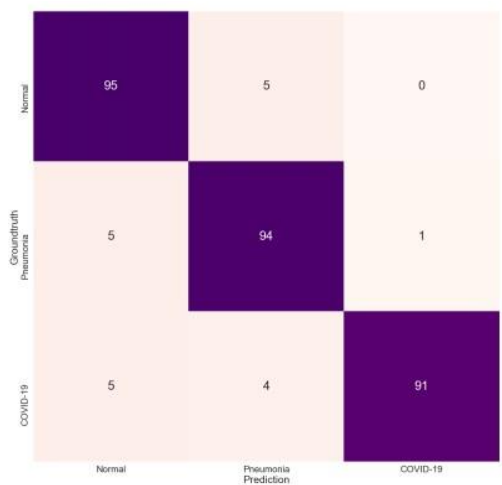

**Figure 14:** Confusion matrix on the 300 test images for COVID-Net. Image from the original COVID-Net paper [18]. COVID-Net results are better than all the open source models that we use for training.

**Confidence Interval:** A confidence interval is a range of values we are fairly sure our true value always lies in. Since the number of images in test set are so few, only 100 images in each class, we see high confidence intervals in most of the cases, both with the open source models, as well as with VisionPro Deep Learning. The best possible way to reduce the confidence interval is to increase the number of images in the test set, by a range, which lies somewhere in the thousands and not in the hundreds. Since the number of COVID-19 images are very few, and we want to make a one-to-one comparison with the results of COVID-Net [18], we use the same number of images provided in the test set of the COVIDx dataset. We calculate a 95% confidence interval on the predicted sensitivity and the positive predicted values, to figure out a possible range of values by which the actual results may vary on the given test data. The confidence interval of the accuracy rates is calculated using the formula:

$$r = z \left( \sqrt{(accuracy(1 - accuracy))} \right) / N$$

where, z is the significance level of the confidence interval (the number of standard deviation of the Gaussian distribution), accuracy is the estimated accuracy (in our cases sensitivity, positive predictive value, and F-score), and N (100 for each class) denotes the number of samples for that class. Here we use 95% confidence interval, for which the corresponding value of z is 1.96 [35].

## 2. Sensitivity

| | | Sensitivity (%) – Results Obtained | | |
|---|---|---|---|---|
| | | **Normal** | **Non-COVID-19** | **COVID-19** |
| 1 | **VGG19** | 93.0 | 80.0 | 80.0 |
| 2 | **ResNet50 V2** | 96.0 | 86.0 | 81.0 |
| 3 | **Densenet121** | **99.0** | 86.0 | 74.0 |
| 4 | **Inception V3** | 97.0 | 85.0 | 92.0 |
| 5 | **COVID-Net** | 95.0 | **94.0** | 91.0 |
| 6 | **VisionPro Deep Learning 1.0 - Entire ROI** | 98.0 | 89.0 | 95.0 |
| 7 | **VisionPro Deep Learning 1.0 - Segmented Lungs** | 98.0 | 91.0 | **97.0** |

**Table 1a:** Sensitivity for each infection type. Best results are highlighted in **BOLD**.

| | | Sensitivity (%) – 95% Confidence Interval | | |
|---|---|---|---|---|
| | | **Normal** | **Non-COVID-19** | **COVID-19** |
| 1 | **VGG19** | 93 ± 5.0 | 80 ± 7.8 | 80 ± 7.8 |
| 2 | **ResNet50 V2** | 96 ± 3.8 | 86 ± 6.8 | 81 ± 7.6 |
| 3 | **Densenet121** | **99 ± 1.9** | 86 ± 6.8 | 74 ± 8.5 |
| 4 | **Inception V3** | 97 ± 3.3 | 85 ± 6.9 | 92 ± 5.3 |
| 5 | **COVID-Net** | 95 ± 4.2 | **94 ± 4.6** | 91 ± 5.6 |
| 6 | **VisionPro Deep Learning 1.0 - Entire ROI** | 98 ± 2.7 | 89 ± 6.1 | 95 ± 4.2 |
| 7 | **VisionPro Deep Learning 1.0 - Segmented Lungs** | 98 ± 2.7 | 91 ± 5.6 | **97 ± 3.3** |

**Table 1b:** Sensitivity calculated with 95% Confidence Interval. Best results are highlighted in **BOLD**.

Sensitivity or Recall measures the true positive rate. It is the proportion of the true positives detected by a model to the total number of positives. The better the sensitivity the better is the model at correctly identifying the infection.

$$Sensitivity = True\ Positive\ / (True\ Positive + False\ Negative)$$

For Normal and COVID-19 classes, VisionPro Deep Learning significantly outperforms all other models, as seen in Table 1. For the Non-COVID-19 class COVID-Net [18] has the best results. VisionPro Deep Learning has a really good sensitivity to COVID-19 images, 95% for the images with the entire ROI selected, and 97% on the images with the lungs segmented. Also, both settings of VisionPro Deep Learning have 98% sensitivity for images belonging to the class Normal.

### 3. Positive Predictive Values

| | Positive Predictive Value (PPV) (%) – Best Results Obtained | | | |
|---|---|---|---|---|
| | | **Normal** | **Non-COVID-19** | **COVID-19** |
| 1 | **VGG19** | 80.0 | 82.0 | 92.0 |
| 2 | **ResNet50 V2** | 86.0 | 83.0 | 94.0 |
| 3 | **Densenet121** | 80.0 | 87.0 | 95.0 |
| 4 | **Inception V3** | 90.0 | 92.0 | 92.0 |
| 5 | **COVID-Net** | 90.5 | 91.3 | **98.9** |
| 6 | **VisionPro Deep Learning 1.0 - Entire ROI** | 89.0 | 95.7 | 96.9 |
| 7 | **VisionPro Deep Learning 1.0 - Segmented Lungs** | **93.3** | **95.8** | 97.0 |

Table 2a: Positive Predictive Value (PPV) for each infection type. Best results are highlighted in **BOLD**.

| | Positive Predictive Value (PPV) (%) – 95% Confidence Interval | | | |
|---|---|---|---|---|
| | | **Normal** | **Non-COVID-19** | **COVID-19** |
| 1 | **VGG19** | 80 ± 7.8 | 82 ± 7.5 | 92 ± 5.3 |
| 2 | **ResNet50 V2** | 86 ± 6.8 | 83 ± 7.3 | 94 ± 4.6 |
| 3 | **Densenet121** | 80 ± 7.8 | 87 ± 6.5 | 95 ± 4.2 |
| 4 | **Inception V3** | 90 ± 5.8 | 92 ± 5.3 | 92 ± 5.3 |
| 5 | **COVID-Net** | 90 ± 5.8 | 91 ± 5.6 | **99 ± 1.9** |
| 6 | **VisionPro Deep Learning 1.0 - Entire ROI** | 89 ± 6.1 | 96 ± 3.8 | 97 ± 3.3 |
| 7 | **VisionPro Deep Learning 1.0 - Segmented Lungs** | **93 ± 5.0** | **96 ± 3.8** | 97 ± 3.3 |

Table 2b: Positive Predictive Value (PPV) calculated with 95% Confidence Interval. Best results are highlighted in **BOLD**.

Positive Predictive value (PPV) or Precision shows the percentage of how many predictions selected the model are relevant.

$$Positive\ Predictive\ Value\ (PPV) = True\ Positive\ /(True\ Positive + False\ Positive)$$

As seen in Table 2, DenseNet121 [16] has the best PPV for Normal images, VisionPro Deep Learning has the best PPV for Non-COVID-19 images and COVID-Net [18] has the best PPV for COVID-19 images. Though it is not the best in comparison, VisionPro Deep Learning still has a high PPV value to COVID-19 images, 96.9% for the images with the entire ROI selected, and 97.0% on the images with the lungs segmented.

### 4. Overall F-scores

F-score takes into consideration both the Sensitivity and PPV of a model. It can be considered as an overall score of the performance of the model.

$$F\ score = True\ Positive\ /(True\ Positive + (1/2 * (False\ Positive + False\ Negative)))$$

As seen in Table 3, out of the open-source architectures InceptionV3 [33] has the best F-score for all the three classes. When compared to InceptionV3, COVID-Net has a higher F-score in the Non-COVID-19 and COVID-19 classes but is slightly less in the Normal Class. VisionPro Deep Learning on the entire image as ROI, outperforms all the open-source architectures and COVID-Net, except in the Non-COVID-19 class where the results are very close, with COVID-Net [18] having an F-score of 92.6% and VisionPro Deep Learning having an F-Score of 92.2%. The setting with segmented lungs, VisionPro Deep Learning outperforms all the open-source models, COVID-Net [18] and even itself with the entire ROI selected. On all classes it has the highest F-score. It gets an F-score of 95.6% on Normal images, 93.3% on Non-COVID-19/Pneumonia images and an F-score of 97.0% on the COVID-19 images. In this setting, VisionPro Deep Learning is only classifying based on the lungs, so there are no artifacts, and the results obtained are highly focused. This helps overcome the black-box idea of Deep Learning results.

|   | F-Score - Results | | | |
|---|---|---|---|---|
|   |   | **Normal** | **Non-COVID-19** | **COVID-19** |
| 1 | **VGG19** | 86.0 | 81.0 | 86.0 |
| 2 | **ResNet50 V2** | 91.0 | 85.0 | 87.0 |
| 3 | **Densenet121** | 88.0 | 86.0 | 84.0 |
| 4 | **Inception V3** | 93.0 | 89.0 | 92.0 |
| 5 | **COVID-Net** | 92.6 | 92.6 | 94.7 |
| 6 | **VisionPro Deep Learning 1.0 - Entire ROI** | 93.8 | 92.2 | 96.0 |
| 7 | **VisionPro Deep Learning 1.0 - Segmented Lungs** | **95.6** | **93.3** | **97.0** |

**Table 3a:** F-Score for each infection type. Best results are highlighted in **BOLD**.

|   | F-Score – 95% confidence Interval | | | |
|---|---|---|---|---|
|   |   | **Normal** | **Non-COVID-19** | **COVID-19** |
| 1 | **VGG19** | 86.0 ± 6.8 | 81.0 ± 7.6 | 86.0 ± 6.8 |
| 2 | **ResNet50 V2** | 91.0 ± 5.6 | 85.0 ± 6.9 | 87.0 ± 6.5 |
| 3 | **Densenet121** | 88.0 ± 6.3 | 86.0 ± 6.8 | 84.0 ± 7.1 |
| 4 | **Inception V3** | 93.0 ± 5.0 | 89.0 ± 6.1 | 92.0 ± 5.3 |
| 5 | **COVID-Net** | 92.6 ± 5.3 | 92.6 ± 5.3 | 94.7 ± 4.6 |
| 6 | **VisionPro Deep Learning 1.0 - Entire ROI** | 93.8 ± 5.0 | 92.2 ± 5.3 | 96.0 ± 3.8 |
| 7 | **VisionPro Deep Learning 1.0 - Segmented Lungs** | **95.6 ± 4.2** | **93.3 ± 5.0** | **97.0 ± 3.3** |

**Table 3b:** F-Score calculated with 95% Confidence Interval. Best results are highlighted in **BOLD.**

VisionPro Deep Learning has the best F-scores on COVID-19 images for both the settings. On the entire ROI it has an F-score of 96.0% and on the segmented lungs it has an F-scores of 97.0%. Overall, on all the three classes, VisionPro Deep Learning achieves an F-score of 94.0% on the entire image as ROI, and an F-score of 95.3% on the segmented lungs. The similarity of the results in both the settings and the heatmaps, both show that even without the lungs being segmented, VisionPro Deep Learning is still predicting its classes based on the actual abnormalities.

As expected, when comparing the confidence intervals, none of the models perform well, due to the significantly low number of images in each class in the test set.

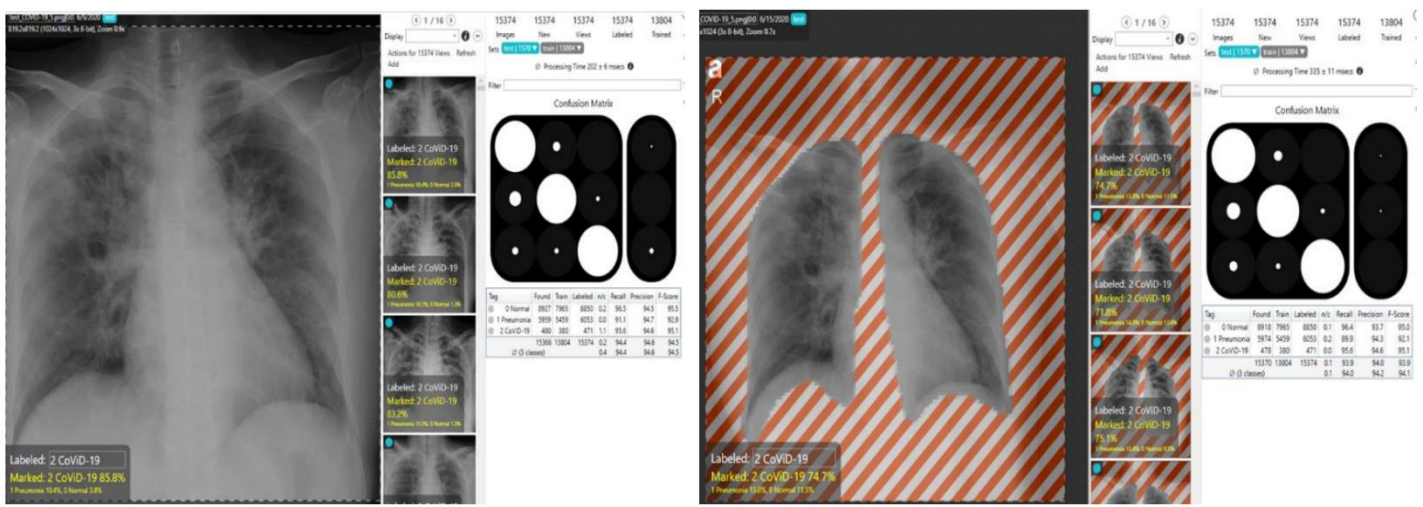

**Figure 15.** VisionPro Deep Learning tested on a previous version of the COVIDx dataset. This dataset has many more images on the test set, in the Normal and Non-COVID-19 classes, but only 91 images on the COVID-19 class. We see the confidence interval improve significantly in classes with higher number of test images.

| Sensitivity (%) – 95% Confidence Interval | | | |
|---|---|---|---|
| | Normal | Non-COVID-19 | COVID-19 |
| VisionPro Deep Learning 1.0 | 97 ± 1.1 | 91 ± 2.3 | 96 ± 4.0 |
| VisionPro Deep Learning 1.0 - Segmented Lungs | 96 ± 1.2 | 90 ± 2.4 | 96 ± 4.0 |

| Positive Predictive Value (%) – 95% Confidence Interval | | | |
|---|---|---|---|
| | Normal | Non-COVID-19 | COVID-19 |
| VisionPro Deep Learning 1.0 | 95 ± 1.4 | 95 ± 1.7 | 95 ± 4.4 |
| VisionPro Deep Learning 1.0 - Segmented Lungs | 94 ± 1.5 | 94 ± 1.9 | 95 ± 4.4 |

| F-score (%) – 95% Confidence Interval | | | |
|---|---|---|---|
| | Normal | Non-COVID-19 | COVID-19 |
| VisionPro Deep Learning 1.0 | 96 ± 1.2 | 93 ± 2.0 | 95 ± 4.4 |
| VisionPro Deep Learning 1.0 - Segmented Lungs | 95 ± 1.4 | 92 ± 2.1 | 95 ± 4.4 |

**Table 4:** Sensitivity, Positive Predictive Value and F-score with 95% confidence interval on the previous COVIDx dataset

We also test VisionPro Deep Learning on a previous version of the COVIDx dataset, which has a total of 15,374 images. They are split into the following number of images in the train set: 7,965 images in the Normal class, 5,459 images in the Non-COVID-19 class and 380 images in the COVID-19 class and the following number of images on the test set of each class: 885 images in the Normal class, 594 images in the Non-COVID-19 class and 91 images in the COVID-19 class. As seen in Figure 15, due to the significantly high number of images in the Normal and Non-COVID-19 class, the confidence interval significantly improves from the previous values ranging from 3%-5% to just 1.0%-2.4%. These results clearly indicate that as the number of test images are increased, the confidence interval improves significantly. Similarly, as this dataset has only 91 images in the COVID-19 class, the confidence interval is hence similar to the previous results.

Also, table 4 indicates that even when the number of images in the test set are significantly increased, the performance of VisionPro Deep Learning does not fall, but rather it still produces sensitivity, PPV and F-scores above 90% in all the classes. If table 4 is compared with the previous results, it is seen that the results are very consistent in VisionPro Deep Learning, even with a change in the number of images in the train and test set. Also, the results of the Sensitivity, PPV and F-scores are very similar with the entire image as ROI and also on the segmented lungs, further indicating that the predictions are made based on the lungs and not on the surrounding artifacts.

**CONCLUSION**

In this study we use COGNEX's Deep Learning Software- VisionPro Deep Learning (version 1.0) and compare its performance with other state of the art Deep Learning architectures. VisionPro Deep Learning has an intuitive GUI making the software very easy to use. Building applications requires no coding skills in any programming language. Little to no preprocessing is required, also decreasing the development time. Imbalanced data is automatically balanced within the software. Once the images are loaded into VisionPro Deep Learning and the right tool is selected, the Deep Learning training can start. After completion of training, it outputs a confusion matrix, along with the various important metrics, such as precision, recall and F-score. Additionally, a report can be generated that identifies all misclassified images. This makes it particularly suitable for radiologists, hospitals, and research workers to harness the power of Deep Learning without advanced coding knowledge. Moreover, as the results from this study indicates, the Deep Learning algorithms in VisionPro Deep Learning are robust and comparable or even better than the various state of the art algorithms available today. The problem of Deep Learning algorithms being a "black box" can be overcome by using a pipeline of tools, stacked sequentially to first segment the lungs then classify only on the segmented lungs. It is like combining a U-Net [17]and Inception [32] model together. This ensures the algorithm does not focus on any artifacts when generating its classification results. A heatmap can be generated to showcase exactly where the model is focusing on while making the predictions. And with both the settings, using the entire image as the Region of Interest and classification on the segmented lungs, VisionPro Deep Learning achieves the highest overall F-scores, surpassing the results of the various open source architectures.

In the future, more testing will be done to understand how changing the number of training images or using augmentations in the training set affects the performance of VisionPro Deep Learning as compared to the other open-source models.

This software is by no means a stand-alone solution in the identification of images of COVID-19 from Chest X-rays, but can aid radiologists and clinicians to achieve a faster and understandable diagnosis using the full potential of Deep Learning, without the prerequisite of having to code in any programming language.

**ACKNOWLEDGEMENT**

We will like to thank COGNEX for providing their latest Deep Learning software for testing, and University of Waterloo, along with Darwin AI for collecting and merging the X-ray images from various sources and for providing the python scripts for generating the COVIDx dataset.

**ADDITIONAL INFORMATION**

Arjun Sarkar, Joerg Vandenhirtz, Jozsef Nagy, David Bacsa, Mitchell Riley are affiliated with COGNEX, which funded the research in this paper.